\documentstyle[12pt]{article}
\makeatletter
\def\maketitle{\par
 \begingroup
 \def\thefootnote{\fnsymbol{footnote}}
 \def\@makefnmark{\hbox
 to 0pt{$^{\@thefnmark}$\hss}}
 \if@twocolumn
 \twocolumn[\@maketitle]
 \else \newpage
 \global\@topnum\z@
 \@maketitle
 \fi
 \thispagestyle{empty}
 \@thanks
 \newpage
 \endgroup
 \setcounter{footnote}{0}
 \let\maketitle\relax
 \let\@maketitle\relax
 \gdef\@thanks{}\gdef\@author{}\gdef\@title{}
 \gdef\@prepri{}\gdef\@address{}
 \let\thanks\relax}
\def\@maketitle{\newpage
 \begin{flushright}
 {\em Ochanomizu University}
 \hfill
 \@prepri \\
 \@date
 \end{flushright}
 \vskip 2em \begin{center}
 {\LARGE \@title \par} \vskip 2em
 {\large \lineskip .5em
 \begin{tabular}[t]{c}\@author
 \end{tabular}\par}
 \vskip .5em
 {\em \begin{tabular}[t]{c}\@address
 \end{tabular}\par}
 \end{center}
 \vskip 1.5em
 \vfill
\@ifundefined{@abst}{}{
 \small
 \begin{center}
 {\bf Abstract\vspace{-.5em}\vspace{0pt}}
 \end{center}
 \begin{quotation} \@abst \end{quotation}
 \vfill
 \gdef\@abst{}}}
\long\def\abst#1{\long\gdef\@abst{#1}}
\def\prepri#1{\gdef\@prepri{OCHA-PP-#1}}
\def\address#1{\gdef\@address{#1}}
\def\section{\@startsection {section}{1}{\z@}{3.5ex plus 1ex minus
 .2ex}{2.3ex plus .2ex}{\Large\bf}}
\def\subsection{\@startsection{subsection}{2}{\z@}{3.25ex plus 1ex minus
 .2ex}{1.5ex plus .2ex}{\large\bf}}
\def\subsubsection{\@startsection{subsubsection}{3}{\z@}{3.25ex plus
 1ex minus .2ex}{1.5ex plus .2ex}{\normalsize\bf}}
\def\eqnarray{%
 \stepcounter{equation}%
 \let\@currentlabel=\theequation
 \global\@eqnswtrue
 \global\@eqcnt\z@
 \tabskip\@centering
 \let\\=\@eqncr
 $$\halign to \displaywidth\bgroup\@eqnsel\hskip\@centering
 $\displaystyle\tabskip\z@{##}$&\global\@eqcnt\@ne
 \hfil$\displaystyle{{}##{}}$\hfil
 &\global\@eqcnt\tw@$\displaystyle\tabskip\z@{##}$\hfil
 \tabskip\@centering&\llap{##}\tabskip\z@\cr}
\makeatother
\def\Sp{{\rm Sp}}
\def\M_g{{\rm M_g}}
\begin{document}
\prepri{32}
\date{December, 1993}
\author{K. Kimura and K. Tesima
%\thanks{also Dep. $\!$of Phys. $\!$Univ. $\!$of Tokyo, Hongo,
%Bunkyo, Tokyo 113, Japan}
}
\address{Department of Physics, Ochanomizu University,
Otsuka, Bunkyo, Tokyo 112, Japan}
\title{Multiplicity with a Thrust Cut}
\abst
{
We evaluate the multiplicity of hadrons in the $e^+e^-$-annihilation
at a given thrust $T$
in the modified leading-log approximation,
including $O(\sqrt{\alpha_s})$ corrections.
The calculation is done at a large value of $\tau =1-T$
by the use of the factorisation which takes place
in the one-particle-inclusive cross section at a given $\tau$.
At a small $\tau$, a different type of factorisation takes place,
which also enable us to evaluate the multiplicity.
Two approaches are compared numerically.
Measuring this quantity near $\tau =1/3$,
we can determine the multiplicity ratio between a gluon-jet and a quark-jet.
}
\maketitle
%%%%%%%%%%%%%%%%%%%%%%%%%%%%%%%%%%%%%%%%%%%%%%%%%%%%%%%%%%%%%%%%%%%%%
\section {Introduction}
The Modified Leading-Log Approximation (MLLA)
\cite{bas}\cite{dok1} with the Local Parton-Hadron Duality
(LPHD)\cite{dok2} has been successful in
describing the majority of the $e^+e^-$-annihilation events
(see, for example, \cite{opa2}\cite{dok3}\cite{smi}).
The shower Monte Carlo programmes based on it \cite{mar}
are widely used for the simulation of experiments.
Nevertheless, our experimental knowledge of the multiple
hadroproduction in {\em gluon jets} is still poor,
or even apparently in contradiction with the theoretical predictions.
The contradiction has been felt particularly in the
multiplicity ratio between a gluon-jet and a quark-jet.

In earlier time, some argued that the multiplicity ratio would be identical
to the ratio between the colour charges of a gluon and a quark \cite{bro}.
Nowadays, this expectation is justified in the following way:
According to LPHD, the average number (multiplicity) of hadrons
produced in a hard process
is proportional to the number of gluons emitted perturbatively
from the hard partons.
Because of the infrared singularity of the emission amplitude,
the majority of the gluons are soft (much less energetic
than the hard parton).
At the leading order in MLLA, the strength of the soft-gluon radiation is
proportional to the colour charge
of the hard parton ($C_A=3$ for a gluon, $C_F=4/3$ for a quark)
with which the jet is associated.
Thus the multiplicity ratio between a gluon-jet and a quark-jet
is expected to be close to $C_A/C_F=9/4$ at high energies.

The ratio has been measured using three-jet events of the
$e^+e^-$-annihilation, identifying one of the jets as a gluon-jet.
The ratio was found much lower than the above expectation
$C_A/C_F$.
For example, OPAL collaboration gave \cite{opa1}
\begin{equation}
r={\langle n\rangle_{\mbox{\tiny g-jet}}\over \langle
n\rangle_{\mbox{\tiny q-jet}}}
=1.267\pm 0.043{\rm (stat.)}\pm 0.055{\rm (syst.)}
\end{equation}
(including neutral particles) and
\begin{equation}
r_{\mbox{\tiny CH}}={\langle n_{\mbox{\tiny CH}}\rangle_{\mbox{\tiny g-jet}}
\over \langle n_{\mbox{\tiny CH}}\rangle_{\mbox{\tiny q-jet}}}=
1.326\pm 0.054{\rm (stat.)}\pm 0.073{\rm (syst.)}
\end{equation}
(charged particles only).
(For earlier experiments, see \cite{rat}.)

The next-to-leading order correction in MLLA
($O(\sqrt{\alpha_s})=O(1/\sqrt{\ln (W^2/\Lambda_{QCD}^2)})$,
$W$: the total energy)
reduces the leading order prediction
$r=9/4$ by about 10 percent \cite{mue1}.
The correction is far too small to explain the discrepancy from the
experimental observation.
(The next-to-next order correction does not change the situation \cite{guf}.)

It should be noted, however, that the multiplicity in a gluon-jet
$\langle n \rangle_{\mbox{\tiny g-jet}}$, which was calculated in MLLA, is
not necessarily identical to the quantity measured in the experiments.
In the theoretical calculation, it is defined as a half of
the total multiplicity from the two hard gluons created by a
gauge invariant gluon source (for example, $F_{\mu \nu}^2$,
where $F_{\mu \nu}$ is the field strength of gluon).

In the experiments, on the other hand, one selects
three-jet events (defined one way or another)
and compares the number of hadrons in each angular region.
At current energies, however, a hard parton
does not necessarily produce a well-collimated jet,
and it may cause biases, depending on how the measurement is made:
If one does not impose the condition that each jet be isolated
(i.e. well-collimated),
the contribution from each hard parton may mix up
one another\footnote{
Strictly speaking, we cannot tell, in a gauge invariant way,
which particle is emitted from which hard parton,
except in the high-energy limit where only collinear particles are produced.
}.
Thus the difference in the multiplicity due to the difference in the
colour charge of the respective hard parton may be reduced in the
comparison of the multiplicity in the respective angular region.
If one selects the events with well-collimated jets, on the other hand,
one is essentially comparing the multiplicity in the direction of
the respective hard parton
(in a narrow angular cone around the jet axis).
Now the stronger radiation of gluons from a hard gluon,
which is responsible for a larger multiplicity in a gluon jet,
also causes a wider angular distribution of hadrons,
owing to the larger recoil from the
{\em multiple} gluon emission from the hard gluon \cite{smi}
(see the angular distribution of hadrons measured in \cite{opa1}).
The multiplicity counted in the direction of the hard gluon, therefore,
is not as large as $C_A/C_F$ times the multiplicity counted
in the direction of the hard quark or antiquark
(see also \cite{dok1}).

Though the magnitude of the biases is in general
hard to estimate theoretically, we may examine it using
the MLLA-based shower Monte Carlo simulation of the experiments.
The simulations done by the experimentalists
(in \cite{opa1} and \cite{rat}) in fact gave reduced multiplicity ratio.
It suggests, at least, that the observed smaller ratio does not
necessarily contradict QCD itself.

In any case, a bias free measurement of the multiplicity ratio
is not easy, because a gauge-invariant two-gluon source
is hard to prepare\footnote{The process
$\gamma\gamma\rightarrow gg$ may be used for this purpose \cite{kho}.}.

In this article, we analyse the multiplicity at a given value of thrust,
where thrust $T$ is defined, in the c.m. $\!$frame, by
\begin{equation}
T = {\rm max}\left\{\frac{\sum_{i} \mid \vec{P}_i\cdot \vec{n}
\mid}{\sum_{i} \mid \vec{P}_i \mid }\right\}~~~~(\vec{n}^2 = 1)
\end{equation}
(the direction of the three-vector $\vec{n}$ is chosen
to maximise the rhs).
$T=1$ would imply that all the particles are parallel
(or antiparallel) to $\vec{n}$.

If $T$ is far from one, it implies that a hard gluon is emitted.
The emitted hard gluon causes an increase of the multiplicity.
At high energies, where most of the events with large $\tau =1-T(<1/3)$
are well-collimated 3-jet events,
the multiplicity increase can be identified with the contribution
from the multiplicity in a gluon-jet.
In this way, we can avoid the experimental bias caused by the
angular restriction, and we can compare the theoretical
calculations with the experimental data without ambiguities.

In Sect.2, we evaluate the multiplicity at a large $\tau$ at the
next-to-leading order of MLLA.
At this order, a hard gluon adds the multiplicity independently
of the quark- and the antiquark-jets.
The multiplicity increase is thus close to $C_A/(2C_F)$ times
the multiplicity from the $q\bar{q}$-pair, which is identical to
the multiplicity $\langle n \rangle_{e^+e^-}$
(without specifying the thrust value)
at the same c.m. $\!$energy as the $q\bar{q}$-pair.

The approximation, however, is not accurate if the energy is not high
enough or $\tau$ is not large enough.
In either case, the gluon jet is not well-isolated, and
higher order corrections cause substantial modifications.

In Sect.3, we evaluate the multiplicity at a small $\tau$,
where no hard gluons are emitted.
At the leading order, the multiple soft-gluon emission
from the $q\bar q$-pair determines
the value of $\tau (\ll 1)$ \cite{tes}\cite{hag}.
Because the soft gluons are emitted
{\em independently} from the hard $q\bar q$-pair,
the one-particle-inclusive amplitude, which gives the multiplicity,
factorises from d$\sigma$/d$\tau$, and the resummation
to all orders can easily be done.

In this approximation, we do not separately evaluate the multiplicity
from a gluon-jet and a quark-jet.
When we apply the approximation to a larger value of $\tau$, however,
we expect two distinctive contributions, one identified
as the multiplicity from the gluon-jet and the other from the
quark- and antiquark-jets.
The approximation is not justified near $\tau =1/3$,
because it assumes that the gluons are emitted independently.
The error, however, would not be large in the intermediate $\tau$ region.
We therefore expect that the prediction in this region is close to the
next-to-leading order prediction obtained in Sect.2,
if the energy is high enough.

We evaluate the multiplicity in the small $\tau$ region making use of the
detailed experimental data of the thrust distribution at 58GeV \cite{top}.
The multiplicity evaluated in this approximation
in the intermediate $\tau$ region ($\tau\approx$ 0.1-0.2)
turns out to be lower than the prediction in Sect.2.
This is because the multiplicity from the hard gluon is not
well-separated from the multiplicity from the $q\bar{q}$-pair.
It suggests that we may have to go to higher energies
in order to be able to determine the multiplicity ratio
from the multiplicity measured at $\tau$ in the intermediate region.

\section {Multiplicity at large $\tau$ (Next-to-Leading Order)}

The one-particle-inclusive (1PI) cross section counts
the number of particles at a given momentum.
When we integrate it over the momentum, and
divide it by the cross section (which counts the number of
the events), we obtain the number of particles per event.
The multiplicity at given $\tau=1-T$ is obtained by dividing
the integrated 1PI cross section at given $\tau$ by the
cross section ${\rm d}\sigma /{\rm d}\tau$.

When $\tau$ is large (but $<1/3$), a hard gluon is emitted.
At high energies, the multiplicity is given as the sum of the two
distinctive contributions:

\noindent (a) The registered particle originates from the hard
$q\bar q$-pair; or

\noindent (b) it originates from the hard gluon.

\noindent In either case, the registered particle is
mostly produced near the direction of the respective hard parton.

At the leading order, the multiplicity at a large $\tau$
is simply $1+C_A/(2C_F)$ times the total multiplicity
$\langle n \rangle_{e^+e^-}$ at the same energy.
The additional multiplicity proportional to $C_A/(2C_F)$
is due to (b) (the multiplicity in a gluon jet).

We first evaluate the next-to-leading order corrections
($O(\sqrt{\alpha_s})$) to (a).
Let $P_1$ ($P_2$) be the momentum of the hard quark (antiquark),
$P_3$ the momentum of the hard gluon, and
$k$ the momentum of the gluon
to which the registered hadron belongs (Fig.1a).
Suppose that the hard gluon is emitted
on the quark side (in the c.m. frame).

At the next-to-leading order, only
the soft and/or collinear emission contributes to the multiplicity.
We can therefore assume $2P_1\cdot P_3\gg m^2$,
where $m^2=2P_1\cdot k$ if the $k$-gluon is emitted on the quark
side, and $m^2=2P_2\cdot k$ if it is emitted on the antiquark side.
The value of thrust is determined by $2P_1\cdot P_3$:
\begin{eqnarray}
1-T&=&\tau =\frac{2P_1\cdot P_3}{W^2}+O\left(\frac{m^2}{W^2}\right)~~,\\
& &2P_1\cdot P_3\gg m^2\;\; .\nonumber
\end{eqnarray}

Owing to the strong inequality,
the cross section for the emission of the two gluons factorises
into the cross section for the hard-gluon emission
(d$\sigma$/d$\tau$) times soft- and/or collinear-gluon emission amplitude.
Therefore, when we divide the 1PI cross section
(illustrated in Fig.1a) by d$\sigma$/d$\tau$, we obtain
\begin{eqnarray}
<n>_{q \bar q}
&=& \int^{\tau W^2/2}_0 \frac{{\rm d}m^2}{m^2}
\int_{m^2/W'^2}^1 {\rm d}x_1
\frac{2C_F}{\pi} \left( \frac{1}{x_1}-1+ \frac{x_1}{2} \right)
\alpha_s (x_1m^2)M_g(x_1m^2)\;,\\
W'^2&=&2P_1\cdot P_2\;,\nonumber
\end{eqnarray}
where $M_g(Q^2)$ is the multiplicity from a gluon emitted
at the transverse momentum $Q$,
and the integration kernel represents the probability of the
emission of a gluon at the longitudinal momentum fraction $x_1$
(cf. $\!$(10) below).
The invariant mass of the $q\bar q$-pair
$W'$ depends on the momentum of the hard gluon $P_3$.
Later, we shall integrate the rhs of (5) over $P_3$
with the weight of the hard-gluon emission probability.

At the next-to-leading order, MLLA gives
the multiplicity function $M_g(Q^2)$ in the form
\cite{bas}\cite{mue2}
\begin{equation}
M_g(Q^2)=C \left( \ln (Q^2/\Lambda^2)\right)^{\gamma_1}
\exp\left[ 2\gamma_0\sqrt{\ln (Q^2/\Lambda_{QCD}^2)}\right]~~,
\end{equation}
where
\begin{eqnarray}
\gamma_0 &=& \sqrt{{C_A\over 2\pi b_0}}~~, \nonumber\\
\gamma_1 &=& -{~1~\over 4} -
{N_f\over 6\pi b_0}\left( 1-{C_F\over C_A}\right)~~, \nonumber\\
b_0 &=& {11C_A-2N_f\over 12\pi } \nonumber
\end{eqnarray}
($N_f$: the number of the active quark flavours).
The normalisation constant $C$ in (6) is not determined
in the perturbation theory, and has to be fixed
by comparing the predictions with the experimental data.
The effective QCD coupling $\alpha_s(Q^2)$ becomes large as $Q$
approaches the QCD mass scale $\Lambda_{QCD}$, and the interaction
becomes nonperturbative.
The emission of the strongly interacting gluons at low $Q$ is
responsible for the bound state formation,
and does not contribute to the multiplicity.
$M_g(Q^2)$, which represents the number of perturbatively emitted gluons,
therefore, should be set zero for $Q^2<Q_0^2$
in the integration on the rhs of (5),
where the infrared cutoff $Q_0$ is of the order of $\Lambda_{QCD}$.
$Q_0$ is an unknown parameter from the point of view of the
perturbation theory.
The dependence on $Q_0$, however, is absorbed in the normalisation
constant $C$ of the multiplicity function $M_g$:
i.e. the ambiguity disappears in the physically observable multiplicities
owing to the renormalisation of $M_g$.

The upper bound of $m^2$ on the rhs of (5) ($m^2<\tau W^2/2$)
is an ambiguous quantity in our approximation.
We have assumed at the accuracy of the next-to-leading order
that $\tau W^2\approx 2P_1\cdot P_3 \gg m^2$.
It implies that the upper bound of $m^2$ can be anywhere below
$\tau W^2$ (if not much smaller than it),
and that the dependence on the upper bound is indeed at the
next-to-next order.
This fact can easily be checked
by the explicit evaluation of the rhs of (5).

The integration on the rhs of (5) gives at the next-to-leading order
\begin{eqnarray}
<n>_{q \bar q}
& = & \frac{4C_F}{C_A} \left\{ 1+ \frac{\gamma_0}{3\sqrt
{\ln W^2/\Lambda_{QCD}^2}} \left( \frac{13}{4}-\frac{N_f}{C_A}
\left(2\frac{C_F}{C_A}-1\right)\right) \right\} M_g(\tau W^2/2)
\nonumber \\
& & -\frac{2C_F}{C_A} \left\{ 1+ \frac{\gamma_0}
{3\sqrt{\ln W^2/\Lambda_{QCD}^2}}\left(\frac{11}{2}-\frac{N_f}{C_A}
\left(2\frac{C_F}{C_A}-1\right) \right)
\right\} M_g\left(\frac{(\tau W^2/2)^2}{W'^2}\right)
\nonumber\\
&=& \frac{2C_F}{C_A} \left\{ 1 +
\frac{\gamma_0}{3\sqrt{\ln W^2/\Lambda_{QCD}^2}}
\left(1-\frac{N_f}{C_A}\left(2\frac{C_F}{C_A}-1\right)\right)
\right\} M_g(W'^2) \;.
\end{eqnarray}

Let us next evaluate the other contribution (b), in which the registered
hadron originates from the hard gluon.
In this case, the multiplicity function $M_g$ is attached
to the hard-gluon line (momentum $P_3$) (Fig.1b).
Other gluons may be emitted from the $q\bar{q}$-pair,
but they have to be soft and/or collinear gluons
at the next-to-leading order in MLLA.
For the unregistered soft and/or collinear emission,
we should note the following:

\noindent $\cdot$ The soft and/or collinear
emission does not change the value of thrust.

\noindent $\cdot$ The soft and/or collinear emission amplitude factorises.

\noindent It implies that the singularity of the real emission amplitude
is cancelled by the singularity of the virtual correction.
Namely, the same cancellation that occurred in the case of the
1PI cross section without specifying the thrust value
takes place also in this case.
We can therefore neglect the emission of the unregistered gluons
at the next-to-leading order.

Suppose that the event consists in only three (on-shell) hard particles:
a quark at a momentum $P_1$, an antiquark at $P_2$, and a gluon at $P_3$.
The $O(\alpha_s)$ cross section for the single gluon emission is given by
\begin{eqnarray}
{\rm d}\sigma &=&\sigma_0
8\pi C_F \alpha_s
\left\{\frac{2P_1\cdot P_3}{(2P_2\cdot P_3)W^2}
+\frac{2P_2\cdot P_3}{(2P_1\cdot P_3)W^2}+\frac{2(2P_1\cdot P_2)}
{(2P_1\cdot P_3)(2P_2\cdot P_3)}\right\}
\frac{{\rm d^3}P_3}{(2\pi)^3 2P_3}\;,\\
&&
W^2=(P_1+P_2+P_3)^2\;,\nonumber
\end{eqnarray}
where $\sigma_0$ is the lowest order cross section
without any gluons emitted ($\gamma^*\rightarrow q\bar{q}$).

We assume for the moment that the antiquark is more energetic
than the quark (in the c.m. frame) ($P_2^0>P_1^0$).
We take the direction of the antiquark as the negative $z$-axis.

The direction of the thrust axis ($\vec{n}$ in (3)) is identical
to the direction of the most energetic parton.
We therefore have to distinguish the two cases.

(i) If the antiquark is more energetic than the gluon
($P_2^0>P_3^0$), the $z$-direction is identical to the thrust axis.
We then have
\begin{eqnarray}
2P_1\cdot P_3&=& \tau W^2 \;,\nonumber\\
2P_2\cdot P_3&=& x(1-\tau )W^2 \;,\nonumber\\
2P_1\cdot P_2 &=& (1-x)(1-\tau )W^2 \;.
\end{eqnarray}
where $x=P_3^+/(P_1+P_3)^+\;(P_i^+=(P_i^0+P_i^3)/\sqrt{2}$).

The differential cross section for the process
$\gamma^*\rightarrow q\bar{q}g$ is in this case
\begin{eqnarray}
\frac{{\rm d}\sigma}{{\rm d}\tau {\rm d}x}
&=& \sigma_0 \frac{C_F\alpha_s}{2\pi}
\left\{\frac{2P_1\cdot P_3}{2P_2\cdot P_3}
+\frac{2P_2\cdot P_3}{2P_1\cdot P_3}
+\frac{2(2P_1\cdot P_2)W^2}{(2P_1\cdot P_3)(2P_2\cdot P_3)}
\right\}\nonumber\\
&=&\sigma_0 \frac{C_F\alpha_s}{2\pi}
\left\{ \frac{\tau}{x(1-\tau )}-\frac{x(1-\tau )}{\tau }
+ \frac{2(1-x)}{\tau x} \right\} \nonumber\\
&\equiv& \sigma_0 \frac{C_F\alpha_s}{2\pi}f_1(\tau ,x)\;.
\end{eqnarray}

The condition $P_2^0>P_3^0$ implies $2P_1\cdot P_3<2P_1\cdot P_2$: namely,
\begin{equation}
x<\frac{1-2\tau}{1-\tau}\;,
\end{equation}
while $P_2^0>P_1^0$ implies $2P_1\cdot P_3<2P_2\cdot P_3$, or
\begin{equation}
\frac{\tau}{1-\tau}<x\;.
\end{equation}

(ii) If $P_2^0<P_3^0$, the direction of the hard gluon is
identical to the thrust axis.
We then have
\begin{eqnarray}
2P_1\cdot P_3&=&\frac{1-x-\tau}{1-x}W^2\nonumber\\
2P_2\cdot P_3&=&\frac{x}{1-x}\tau W^2\;,\nonumber\\
2P_1\cdot P_2&=&\tau W^2\;,
\end{eqnarray}
and
\begin{eqnarray}
\frac{{\rm d}\sigma}{{\rm d}\tau{\rm d}x}&=&\frac{\sigma_0}{1-x}
\frac{C_F\alpha_s}{2\pi} \left\{\frac{2P_1\cdot P_3}{2P_2\cdot P_3}
+\frac{2P_2\cdot P_3}{2P_1\cdot P_3}
+\frac{2(2P_1\cdot P_2)W^2}{(2P_1\cdot P_3)(2P_2\cdot P_3)}
\right\}\nonumber\\
&=&\frac{\sigma_0}{1-x}
\frac{C_F\alpha_s}{2\pi}\left\{ \frac{x\tau }{1-x-\tau}+
\frac{1-x-\tau }{x\tau }+2\frac{(1-x)^2}{x(1-x-\tau )}
\right\} \nonumber\\
&\equiv& \sigma_0 \frac{C_F\alpha_s}{2\pi}f_2(\tau ,x)
\end{eqnarray}

The condition $P_2^0<P_3^0$ implies
$2P_1\cdot P_3>2P_1\cdot P_2$: namely,
\begin{equation}
x< \frac{1-2\tau }{1-\tau }\; ,\\
\end{equation}
while $P_2^0>P_1^0$ now implies
\begin{equation}
\frac{1-\tau }{1+\tau }<x \\
\end{equation}

Finally, the case $P_1^0>P_2^0$ gives the same cross section (now the
direction of the quark is taken as the negative $z$-direction),
and thus simply doubles the contribution.

When a gluon is emitted {\it at a large angle},
the angular ordering of the succeeding soft-gluon emission
is not exact at the next-to-leading order.
Accordingly, the multiplicity from the $P_3$-gluon is not
simply identical to $M_g(Q^2)$, where $Q^2$ is the transverse
component square of $P_3$.
Instead, it is given by
$M_g(2P_1\cdot P_3)+M_g(2P_2\cdot P_3)-M_g(2P_1\cdot P_2)$\footnote{
In order for this expression to be correct, $P_3$ has to be hard}.
The multiplicity from the hard gluon $\langle n\rangle_g$
at a given thrust value ($0\ll \tau <1/3$) is then obtained by
averaging it with the cross sections (10) and (14)
over the range of $x$ given by (11),(12) and (15),(16) respectively:
\begin{eqnarray}
<n>_g
&=&\int {\rm d}x \left\{ M_g(2P_1 \cdot P_3)
+M_g(2P_2 \cdot P_3)-M_g(2P_1 \cdot P_2)\right\}
\frac{{\rm d}\sigma/({\rm d}x{\rm d}\tau )}{{\rm d}\sigma/{\rm d}\tau}
\nonumber \\
&=& \frac{1}{A(\tau)}
\int_{{\tau \over 1-\tau}}^{{1-2\tau \over 1-\tau}} {\rm d}x
\left\{ M_g(\tau W^2 )+M_g\left( x(1-\tau) W^2 \right)-
M_g\left( (1-x)(1-\tau)W^2\right)\right\}
\nonumber \\
& & \times f_1(\tau ,x)\nonumber \\
&+&{1 \over A(\tau)}\int_{1-\tau \over 1+\tau}^{1-2\tau \over 1-\tau}
{\rm d}x \left\{M_g\left( {1-x-\tau \over 1-x}W^2 \right)
+M_g\left( {x \over 1-x}\tau W^2 \right)
-M_g(\tau W^2) \right\}
\nonumber\\
& & \times f_2(\tau ,x)~,
\end{eqnarray}
where
\begin{eqnarray}
A(\tau) &=& F_1(\tau )+F_2(\tau )~, \nonumber\\
F_1(\tau) &=& \int_{{\tau \over 1-\tau}}^{{1-2\tau \over 1-\tau}}
{\rm d}x f_1(\tau ,x)
\nonumber\\
&=&{3\tau ^2+8\tau -3 \over 2\tau (1-\tau )}+{\tau ^2-2\tau
+2 \over \tau (1-\tau )}\ln {1-2\tau \over \tau}\;,
\nonumber\\
F_2(\tau ) &=& \int_{{1-\tau \over 1+\tau}}^{{1-2\tau \over 1-\tau}}
{\rm d}x f_2(\tau ,x)
\nonumber\\
&=&{1+\tau^2 \over 1-\tau}\left\{ \ln {(1-\tau )^2 \over \tau (1+\tau )}
-{1 \over \tau}\ln{(1-\tau )^2\over (1-2\tau )(1+\tau )} \right\}
+2\ln {1+\tau \over 2(1-\tau )}~.
\end{eqnarray}
$f_1$ and $f_2$ are defined in (10) and (14).

Let us now go back to $\langle n\rangle_{q\bar q}$.
Averaging the rhs of (7) over the hard-gluon momentum,
we obtain
\begin{eqnarray}
<n>_{q \bar q}
&=& \frac{2C_F}{C_A} \left\{ 1 +
\frac{\gamma_0}{3\sqrt{\ln W^2/\Lambda_{QCD}^2}}
\left(1-\frac{N_f}{C_A}\left(2\frac{C_F}{C_A}-1\right)\right) \right\}
\int {\rm d}x M(W'^2)
\frac{{\rm d}\sigma/({\rm d}x{\rm d}\tau )}{{\rm d}\sigma/{\rm d}\tau}
\nonumber \\
&=& \frac{2C_F}{C_A} \left\{ 1 +
\frac{\gamma_0}{3\sqrt{\ln W^2/\Lambda_{QCD}^2}}
\left(1-\frac{N_f}{C_A}\left(2\frac{C_F}{C_A}-1\right)\right) \right\}
\frac{1}{A(\tau)}
\nonumber\\
&\times &\left\{
\int_{{\tau \over 1-\tau}}^{{1-2\tau \over 1-\tau}} {\rm d}x
M_g\left( (1-x)(1-\tau)W^2\right) f_1(\tau ,x)
+ \int_{1-\tau \over 1+\tau}^{1-2\tau \over 1-\tau} {\rm d}x
M_g(\tau W^2) f_2(\tau ,x)\right\}\, .\nonumber\\
\end{eqnarray}

The multiplicity at a given $\tau$ is the sum of the two contributions:
\begin{equation}
<n>_{\tau} =<n>_{q\bar{q}}+<n>_g
\end{equation}

The thrust-dependent multiplicity ratio $R(\tau)$ is defined by
\begin{equation}
R(\tau)={<n>_{g} \over <n>_{q}}~~,
\end{equation}
where
\begin{equation}
<n>_{q} \equiv
{1 \over 2} <n>_{q \bar{q}}~~.
\end{equation}

The expressions (17) and (19) can be expanded around $\tau =1/3$.
At $O(\sqrt{\alpha_s})$, we obtain
\begin{eqnarray}
<n>_g & = &
\left\{ 1+ \frac{\gamma_0}{\sqrt{\ln (W^2/\Lambda_{QCD}^2)}}
\frac{B_1(\tau)}{A(\tau)}\right\} M_g\left( {W^2 \over 3}\right)~,\\
<n>_{q \bar q} & = &
\frac{2C_F}{C_A} \left\{ 1 +
\frac{\gamma_0}{3\sqrt{\ln W^2/\Lambda_{QCD}^2}}
\left(1-\frac{N_f}{C_A}\left(2\frac{C_F}{C_A}-1\right)\right) \right\}
\nonumber\\
&&\times
\left\{ 1+ \frac{\gamma_0}{\sqrt{\ln (W^2/\Lambda_{QCD}^2)}}
\frac{B_2(\tau)}{A(\tau)}\right\} M_g\left( {W^2 \over 3}\right)~,
\end{eqnarray}
\begin{eqnarray}
B_1(\tau)&=&\int_{{\tau \over 1-\tau}}^{{1-2\tau \over 1-\tau}}
{\rm d}x f_1(\tau ,x) \ln {3 \tau x \over 1-x} \nonumber \\
&+& \int_{{\tau \over 1-\tau}}^{{1-2\tau \over 1-\tau}}
{\rm d}x f_2(\tau ,x)
\ln {3x(1-x- \tau ) \over (1-x)^2}~, \\
B_2(\tau)&=&\int_{{\tau \over 1-\tau}}^{{1-2\tau \over 1-\tau}}
{\rm d}x f_1(\tau ,x)\ln \left( 3(1- \tau)(1-x)\right) \nonumber \\
&+& \int_{{\tau \over 1-\tau}}^{{1-2\tau \over 1-\tau}}
{\rm d}x f_2(\tau ,x)\ln (3 \tau)
\end{eqnarray}
The ratio $R(\tau )$ is thus
\begin{eqnarray}
R(\tau )&=& \frac{C_A}{C_F}
\left\{ 1+ \frac{\gamma_0}{\sqrt{\ln (W^2/\Lambda_{QCD}^2)}}
\frac{B_1(\tau)}{A(\tau)}\right\}
\left\{ 1+ \frac{\gamma_0}{\sqrt{\ln (W^2/\Lambda_{QCD}^2)}}
\frac{B_2(\tau)}{A(\tau)}\right\}^{-1}
\nonumber\\
&&\times \left\{ 1+\frac{\gamma_0}{3\sqrt{\ln W^2/\Lambda_{QCD}^2}}
\left(1-\frac{N_f}{C_A}\left(2\frac{C_F}{C_A}-1\right)\right) \right\}^{-1}
\end{eqnarray}

We can easily perform the integral on the rhs of (25) and (26) to obtain
\begin{eqnarray}
B_1(\tau ) &=& \ln (3\tau )F_1(\tau )-F_3(\tau )+{1 \over 2}F_5(\tau )
+\ln 3~F_2(\tau )\nonumber \\
& & -F_4(\tau )+F_6(\tau )-F_7(\tau )~~, \\
B_2(\tau ) &=& \ln \left( 3(1-\tau )\right) F_1(\tau )
-{1 \over 2}F_5(\tau )+\ln (3\tau )F_2(\tau )~,
\end{eqnarray}
where
\begin{eqnarray}
F_3(\tau )
&=&
\int_{{\tau \over 1-\tau}}^{{1-2\tau \over 1-\tau}} {\rm d}x
f_1(\tau ,x) \ln {1 \over x}
\nonumber\\
&=&
{1 \over \tau (1-\tau )}\left\{ {1 \over 2}(1-2\tau )(3+2\tau )
\ln {1-2\tau \over 1-\tau}
+{\tau \over 2}(\tau -4)\ln {\tau \over 1-\tau }
\right. \nonumber\\
& &\left. +{1 \over 4}(3\tau -1)(7+\tau )
-{1 \over 2}(\tau^2 -2\tau +2)\ln {(1-2\tau )\tau \over
(1-\tau )^2}\ln {1-2\tau \over \tau} \right\}
\nonumber
\end{eqnarray}
\begin{eqnarray}
F_4(\tau )
&=&
\int_{1-\tau \over 1+\tau}^{1-2\tau \over 1-\tau}
{\rm d}x f_2(\tau ,x)
\ln {1 \over x}
\nonumber\\
&=&
-{1+\tau^2 \over 1-\tau }
\left\{ \Sp \left( {\tau^2 \over (1-\tau)^2} \right)
- \Sp \left( {\tau \over 1+\tau } \right)
\right\}+2\left\{ \Sp \left( {\tau \over 1-\tau }\right)
- \Sp \left( {2\tau \over 1+\tau } \right)\right\}
\nonumber\\
& &
+{1+\tau ^2 \over 1-\tau}\ln (1-\tau )\ln
{\tau (1+\tau ) \over (1-\tau )^2}
-{1+\tau^2 \over 2\tau (1-\tau )}\ln {1-2\tau \over
1+\tau }\ln {(1-2\tau )(1+\tau ) \over
(1-\tau )^2}
\nonumber
\end{eqnarray}
\begin{eqnarray}
F_5(\tau )
&=&
\int_{{\tau \over 1-\tau}}^{{1-2\tau \over 1-\tau}} {\rm d}x
f_1(\tau ,x)
\ln {1 \over (1-x)^2}
\nonumber\\
&=&
-{2+3\tau \over 1-\tau }\ln {\tau \over 1-\tau}
+{3-6\tau \over \tau (1-\tau )}\ln {1-2\tau \over 1-\tau }
-{(1-3\tau )(5+3\tau ) \over 2\tau (1-\tau )}
\nonumber\\
& &
+{2(\tau ^2-2\tau +2) \over \tau (1-\tau )}
\left\{ \Sp \left( {1-2\tau \over 1-\tau }\right)
- \Sp \left( {\tau \over 1-\tau }\right) \right\}
\nonumber
\end{eqnarray}
\begin{eqnarray}
F_6(\tau )
&=&
\int_{1-\tau \over 1+\tau}^{1-2\tau \over 1-\tau}
{\rm d}x f_2(\tau ,x)
\ln {1 \over (1-x)^2}
\nonumber\\
&=&
2{1+\tau ^2 \over 1-\tau }\left\{ \Sp (\tau )
- \Sp \left( {1-\tau \over 2}\right)
+ {1 \over 2}\ln {2(1-\tau ) \over 1+\tau }
\ln {1-\tau ^2 \over 2}
+\ln \tau \ln {\tau (1+\tau ) \over (1-\tau )^2}\right\}
\nonumber\\
& &
-2\ln {2\tau ^2 \over 1-\tau ^2}
\ln {1+\tau \over 2(1-\tau )}
+{2(1+\tau ^2) \over \tau (1-\tau )}
\left\{ \Sp \left( {1-2\tau \over 1-\tau }\right)
- \Sp \left( {1-\tau \over 1+\tau }\right) \right\}
\nonumber
\end{eqnarray}
\begin{eqnarray}
F_7(\tau )
&=&
\int_{1-\tau \over 1+\tau}^{1-2\tau \over 1-\tau}
{\rm d}x f_2(\tau ,x)
\ln {1 \over 1-x-\tau }
\nonumber\\
&=&
{1+\tau ^2 \over 2(1-\tau )}\ln {\tau ^3 \over 1+\tau }
\ln {\tau (1+\tau ) \over (1-\tau )^2}
+2\left\{ \Sp \left( {1+\tau \over 2}\right) -\Sp (1-\tau )
\right\}
\nonumber\\
& &
+\ln {1-\tau ^2 \over 2\tau ^2} \ln {1+\tau \over 2(1-\tau )}
+ {1+\tau ^2 \over \tau (1-\tau )}
\left\{ \Sp \left( {1-2\tau \over (1-\tau )^2} \right)
-\Sp \left( {1 \over 1+\tau }\right) \right\}
\nonumber\\
& &
-{1+\tau ^2 \over \tau (1-\tau )}
\ln (1-\tau )\ln {(1-2\tau )(1+\tau ) \over (1-\tau )^2}~~.
\end{eqnarray}
$\Sp (x)$ is the Spence function defined by
\begin{equation}
\Sp (x)=\int^x_0 {{\rm d}x\over x}\, \ln {1\over 1-x}~~.
\end{equation}

In the limit $\tau\rightarrow 1/3$, we find that
$B_i(\tau )/A(\tau )\rightarrow 0,\; (i=1,2)$, and
\begin{equation}
R(1/3)=r\;,
\end{equation}
where $r$ is the conventional multiplicity ratio
between a quark-jet and a gluon-jet
calculated at the next-to-leading order \cite{mue1}:
\begin{equation}
r=\frac{C_A}{C_F}\left\{1-\frac{\gamma_0}{3\sqrt
{\ln W^2/\Lambda_{QCD}^2}} \left( 1-\frac{N_f}{C_A}
(2\frac{C_F}{C_A}-1)\right) \right\}
\end{equation}

We show in Fig.2 the prediction for $\langle n\rangle_{\tau}$ divided
by $\langle n\rangle_{e^+e^-}$ at $W$=58GeV.
$\langle n\rangle_{e^+e^-}$ is given by
\begin{eqnarray}
<n>_{e^+e^-} &=&
\int^{W^2}_0 \frac{{\rm d}m^2}{m^2}
\int_{m^2/W^2}^1 {\rm d}x_1
\frac{2C_F}{\pi} \left( \frac{1}{x_1}-1+ \frac{x_1}{2} \right)
\alpha_s (x_1m^2)M_g(x_1m^2)
\nonumber\\
&=& \frac{2C_F}{C_A} \left\{ 1 +
\frac{\gamma_0}{3\sqrt{\ln W^2/\Lambda_{QCD}^2}}
\left(1-\frac{N_f}{C_A}\left(2\frac{C_F}{C_A}-1\right)\right)
\right\} M_g(W^2) \;.
\end{eqnarray}
In Table 1, the prediction at 58GeV is compared numerically with
that at 91GeV.
We find that the differences are negligibly small.

In Fig.2, we also show $R(\tau )$ at $W$=58GeV.
It reduces appreciably as $\tau$ decreases, while
change in $\langle n\rangle_{\tau}$ is more moderate.
As the emitted gluon becomes harder (at larger $\tau$),
it carries away a portion of energy from the $q\bar q$-pair,
and the increase in $\langle n\rangle_g$ is partially compensated
by the decrease in $\langle n\rangle_{q\bar q}$ in their sum
$\langle n\rangle_{\tau}$ (recoil effect).

The multiplicity ratio $R(\tau)$ is in general {\it not}
a directly measurable quantity.
What we measure without ambiguity is not
$\langle n\rangle_{g}$ or $\langle n\rangle_{q}$ itself,
but their sum $\langle n\rangle_{\tau}$.
However, $R(1/3)=r$ can be determined experimentally in the following way:
First, we measure $\langle n\rangle_{\tau}$ at $\tau=1/3$.
(In practice,
because of the lack of statistics near $\tau =1/3$,
we measure $\langle n\rangle_{\tau}$ for $\tau<1/3$,
and extrapolate it to $\tau=1/3$, assuming a smooth dependence on $\tau$.)
Now we note that $\langle n\rangle_{q\bar q}$ at $\tau =1/3$
is identical to $\langle n\rangle_{e^+e^-}$ at the
c.m.$\!$ energy $W/\sqrt{3}$,
which is the c.m.$\!$ energy of the $q\bar q$ pair (compare (7) with (34)).
We can therefore use the experimentally measured
$\langle n\rangle_{e^+e^-}$ at the c.m.$\!$ energy $W/\sqrt{3}$
for $\langle n\rangle_{q\bar q}$ at $\tau =1/3$.
Then $\langle n\rangle_{g}$ is defined as $\langle n\rangle_{\tau =1/3}$
minus $\langle n\rangle_{q\bar q}$, and $R(1/3)$ is obtained by (21).

It should be noted that the above result (23) and (24)
cannot be applied to $\tau>1/3$, because the $O(\alpha_s)$
cross section vanishes for $\tau>1/3$.

\section{Small $\tau$ region}

So far, we have analysed the large $\tau$ region, assuming that
the event consists in three well-separated jets (i.e. the invariant
mass of each jet is much smaller than $W$).
When $\tau$ is small, multiple soft-gluon emission determines
the value of $\tau$ \cite{tes},
and it is no more possible to isolate a single gluon jet.
In this case, however, the soft gluons can be regarded
as emitted independently from the initial hard $q\bar{q}$-pair,
and the soft-gluon emission amplitude factorises
(as is the case with soft-photon emission in QED).
Making use of the factorisation, we can evaluate
the multiplicity at a given $\tau$ without dividing it into
the contributions from isolated jets.

Let us first remind ourselves that if all the emitted gluons are soft
$\tau$ can be expressed as the sum
of the contribution of each gluon \cite{tes}:
\begin{equation}
\tau\approx\sum_i m_i^2\: ,\:\:\:
m_i^2={\rm min}\{2P_1\cdot k_i, 2P_2\cdot k_i\}\: ,
\end{equation}
where $k_i$ is the momentum of each soft gluon.
The emission amplitude of $n$ soft gluons is
factorised into the $n$ identical amplitudes of
one-gluon emission (Eikonal Approximation),
and we obtain the thrust distribution $F(\tau)$ in the form of a series
\begin{eqnarray}
F(\tau)&=&{1\over \sigma}{\rm{d}\sigma\over \rm{d}\tau} \nonumber\\
&=& \Delta(W^2,Q_0^2) \left\{ \Gamma (\tau W^2)
+\frac{1}{2!}\int^{\tau}_0 {\rm d}\tau_1 \Gamma (\tau_1W^2)
\Gamma \left( (1-\tau_1) W^2\right)
+ \cdot \cdot \cdot \right\}\nonumber\\
& &\hspace{2.0cm}(\tau\neq 0)\;\;,
\end{eqnarray}
where $\Gamma(\tau W^2)$ is the factorised amplitude
for the single soft-gluon emission (cf. $\!$(10))
\begin{equation}
\Gamma(\tau W^2)=\frac{2C_F}{\pi} \frac{1}{\tau} \int ^1 _{\tau}
\frac{{\rm d}x}{x} \alpha_s(x\tau W^2)
\theta(x\tau W^2-Q_0^2)\;\;,
\end{equation}
and $\Delta$ is the normalisation factor:
\begin{equation}
\Delta(W^2,Q_0^2)= \exp \left[ -\int ^{1/3}_0 d\tau
\Gamma(\tau W^2) \right]\;\;.
\end{equation}
$Q_0$ is the lower bound of the transverse momentum of
the emitted gluon, below which the soft gluon
does not contribute to the particle proliferation.

The multiplicity at given $\tau$ is equal to the
1PI cross section at given $\tau$ integrated over
the whole phase space and divided by
${\rm d}\sigma/{\rm d}\tau$.
The integrated 1PI cross section
at a given $\tau$ (divided by $\sigma$) is obtained by substituting
the multiplicity $M_g(x\tau W^2)$ for
one of the $\Gamma$s in each term of the series on the rhs of (36),
thereby substituting $1/(j-1)!$ for the Bose factor $1/j!$
in the $j$-th term:
\begin{eqnarray}
F(\tau )<n>_{\tau}
&=&\frac{1}{\sigma}\frac{{\rm d}\sigma}{{\rm d}\tau}<n>_{\tau}
\nonumber\\
&=& \Delta(W^2,Q_0^2) \left\{ M'(\tau W^2)+\int^{\tau}_0
\frac{{\rm d}\tau_1}{\tau_1} M'(\tau_1 W^2)
\Gamma\left( (1-\tau_1)\right) W^2) \right.
\nonumber\\
& & \left. +\frac{1}{2!} \int^{\tau}_0 \frac{{\rm d}\tau_1}{\tau_1}
\int^{\tau-\tau_1}_0{\rm d}\tau_2
M'(\tau_1 W^2) \Gamma (\tau_2 W^2) \Gamma\left( (1-\tau_1-\tau_2) W^2\right)
+ \cdot \cdot \cdot \right\} \nonumber\\
&=& \int ^\tau_0 \frac{{\rm d}\tau_1}{\tau_1}
2M'(\tau_1 W^2) \left\{ \delta (\tau_1-\tau)
+\Gamma\left((\tau-\tau_1)W^2\right) \right. \nonumber\\
& & \left. + \frac{1}{2!} \int^{\tau-\tau_1}_0 {\rm d}\tau_2
\Gamma(\tau_2 W^2)
\Gamma\left((1-\tau_1-\tau_2) W^2\right) +\cdot \cdot \cdot \right\}\;\;,
\end{eqnarray}
where
\begin{equation}
M'(\tau W^2)=\frac{2C_F}{\pi} \int_{\tau}^{1} \frac{{\rm d}x}{x}
\alpha_s (x\tau W^2) M_g(x\tau W^2) \;\;.
\end{equation}
Comparing the rhs of (39) with (36), we obtain
\begin{equation}
<n>_{\tau}=
\frac{\int_0^{\tau} \frac{{\rm d}\tau_1}{\tau_1}
M'(\tau_1 W^2) F(\tau-\tau_1)}{F(\tau )}\;\;.
\end{equation}

In (37) and (40), we included only the leading order term.
The next-to-leading order corrections can be included
by substituting
\begin{eqnarray}
\Gamma(\tau W^2)&=& \frac{1}{\tau} \int ^1 _{\tau}{\rm d}x\left\{C_F
\frac{\alpha_s(x\tau W^2)}{\pi}\left( \frac{1}{x}-1+\frac{x}{2}\right)
+a^{(2)}\left(\frac{\alpha_s(x\tau W^2)}{\pi}\right)^2
\frac{1}{x}\right\}\nonumber\\
& &\;\;\;\times\theta(x\tau W^2-Q_0^2)\;\;,\\
a^{(2)}&=&\frac{C_F}{2}\left(\frac{67}{6}-\frac{\pi^2}{2}-\frac{5N_f}{9}
\right) \;\;,\nonumber
\end{eqnarray}
for (37), with two-loop $\alpha_s$,
$\Lambda_{QCD}=\Lambda_{\overline{MS}}$; and
\begin{equation}
M'(\tau W^2)=\frac{2C_F}{\pi} \int_{\tau}^{1} {\rm d}x
\left(\frac{1}{x} -1+\frac{x}{2} \right)
\alpha_s (x\tau W^2) M_g(x\tau W^2)
\end{equation}
for (40)\footnote{The term proportional to
$a^{(2)}$ in the gluon emission amplitude off a quark (see (42))
gives the next-to-next correction in evaluating $M'(\tau W^2)$.
In fact, the leading-log approximation (36) is organised differently
from MLLA, and we cannot tell a priori whether the next-to-next
corrections in MLLA are in fact smaller than the next-to-leading
order corrections in (36). (The former haven't been calculated yet.)}.

The distribution $F(\tau)$ can be evaluated either directly (numerically)
\cite{tes}\cite{hag}, or using Laplace transformation \cite{cat2}.
It is an infrared-safe quantity in the following sense:
Though each term of the series (36) depends on
the infrared cutoff $Q_0$, the dependence disappears
in the sum at the high energy limit.
In fact, the result somewhat depends on the value of $Q_0$
at a finite energy, which
gives rise to the ambiguity of the order of $Q_0/W$.
We do not know theoretically the exact value of $Q_0$,
because it is associated with the non-perturbative hadronisation process.

In order to minimise the theoretical ambiguity,
we use the experimentally observed thrust distribution
for $F(\tau )$ on the rhs of (41).
Detailed data in the small $\tau$ region was obtained
by TOPAZ collaboration at TRISTAN at $W$=58GeV \cite{top}.
The result of the numerical evaluation of (41) is shown in Fig.3.
Note that the experimental data of $F(\tau )$
are given as the average values in finite bin sizes.
Though the numerator of the rhs of (41) is continuous in $\tau$,
its denominator is not.
In order to reduce the error,
we evaluate (41) only at the middle value of $\tau$ in each bin,
where the data of $F(\tau )$ is not far from its real value.
In the integrand on the numerator, $F(\tau-\tau_1)$
is regarded as a step function (constant in each bin).
In Fig.3, the result at each $\tau$ is connected
one another by straight lines.
The curve thus obtained is still not smooth because of the inaccuracy
included in this procedure.

Let us now examine the validity of the approximation.
First, we discuss the small $\tau$ region.

It has sometimes been argued that at the current energies
the thrust distribution in the small
$\tau$ region ($\tau<0.07$ or so) was determined mostly
by the hadronisation process.
The argument was partly based on the observation in the Monte Carlo
simulation studies:
The thrust distribution at the parton level (with partons
produced above certain transverse momentum cutoff of the order of 1GeV)
is modified substantially in the small $\tau$ region by the hadronisation
(i.e. with the phenomenological hadronisation models
to convert the particles produced by the QCD parton shower into hadrons).
If so, use of (36) for $F(\tau )$ (also (39) for $\langle n\rangle_{\tau}$)
would not be justified in the small $\tau$ region.

A recent study \cite{hag}, however, has shown that when we lower the
cutoff in the partonic thrust distribution, the result reproduces
the data quite well.
It suggests either that the non-perturbative hadronisation process
is almost irrelevant to the distribution,
or that the non-perturbative hadroproduction can be
simulated by the perturbative soft-gluon emission.
Whichever the case, we may safely assume
that the LPHD holds also in small $\tau$-processes\footnote{
The assumption that the LPHD is correct with a small infrared cutoff
is consistent with the experiences in the particle spectrum\cite{opa2}.}.

Another question to be addressed is on the recoil effect.
In deriving the formulae, we have neglected the fact that
an emitted gluon carries out a part of the energy of the
parent quark, and accordingly the succeeding emission makes
a reduced contribution to $\tau$ (recoil effect) \cite{hag}.
The recoil is large at the large $\tau$ region.
At the small $\tau$ region, where soft-gluon emission dominates,
the recoil is negligible.
In fact, the correction at the large $\tau$ region
modifies the distribution at a small
$\tau$ through the change in its overall normalisation.
The change in the normalisation of $F(\tau )$, however, is canceled
on the rhs of (41) between the numerator and the denominator.

At large $\tau$, on the other hand, (41) overestimates the multiplicity.
The thrust distribution in the large $\tau$-region is determined
by the Feynmann diagrammes at the lowest orders in $\alpha_s$.
Because the $O(\alpha_s)$ matrix element vanishes for $\tau\geq 1/3$,
the thrust distribution decreases rapidly above $\tau\approx 1/3$.
The numerator on rhs of (41), on the other hand,
does not vanish at $\tau\geq 1/3$ even if $F(\tau_1)$
vanishes for $\tau_1\geq 1/3$ in its integrand.
The overestimation is particularly large when $\tau$ is close to 1/3.

Apart from this obvious overestimation near $\tau=1/3$
(see Fig.3 at $\tau\approx 1/3$), the recoil effect
causes the reduction of the multiplicity in a wider $\tau$ region.
In addition to the recoil effect on $F(\tau)$,
already discussed above, the recoil from the emission of a hard gluon
reduces the multiplicity from the quark-antiquark pair, which partly
compensates the multiplicity increase from the hard gluon
(see the discussion given at the end of Sct.2).
This effect is not included in (41).
With the recoil effect,
the multiplicity increase in $\tau$ at large $\tau$
would be less than the one estimated by (41).

For the intermediate value of $\tau$ ($\tau\approx$0.1-0.2),
the overestimation reduces.
On the other hand, we note that at very high energies
the integrand in the numerator on the rhs of (41) has two sharp
peaks, one at $\tau_1\approx 0$ (near the lower bound)
and the other at $\tau_1\approx \tau$ (near the upper bound).
The contribution from the former peak corresponds to the
multiplicity in the quark-(or antiquark-)jet,
while the contribution from the latter corresponds to the
multiplicity in the gluon-jet.
We may therefore expect that (41) would give a prediction
close to (20) in the intermediate $\tau$ region.

In Fig.3, we also show $\langle n\rangle_{\tau}$ evaluated by (20)
in order to compare it with (41).
We find that (41) gives smaller multiplicity than (20)
(except near $\tau =1/3$),
despite the fact that the neglect of the recoil effect
tends to give an overestimation in (41)
in the large and intermediate $\tau$ region.
The reason for the difference is not difficult to find.
At 58GeV, the thrust distribution is not very sharply peaked at
very small $\tau$,
so that in the integration in the numerator on the rhs of (41)
the peak corresponding to the multiplicity in the gluon jet does not
develop enough to give a contribution separate from the multiplicity
from the $q\bar q$-pair.
(The maximum of d$\sigma$/dln$(1/\tau)$ is at $\tau\approx 0.043$.)
If the jets are not collimated enough,
it may be difficult to determine the multiplicity ratio
from the multiplicity data with a thrust cut at this energy.

Finally, let us make a comment on the determination of the normalisation
of the multiplicity function $M_g$, which has been implicit so far.
Because $F(\tau )$ is a normalised distribution,
we should have
\begin{equation}
\int_0^{1/2} {\rm d}\tau F(\tau )<n>_{\tau}=<n>_{e^+e^-}\; .
\end{equation}
On the other hand, when we integrate $F(\tau )\langle n\rangle_{\tau}$
given by (41) over $0<\tau<1$,
we obtain
\begin{eqnarray}
\int^1_0{\rm d}\tau F(\tau )<n>_{\tau}
&=&
\int^1_0{\rm d}\tau \int_0^{\tau} \frac{{\rm d}\tau_1}{\tau_1}
M'(\tau_1 W^2) F(\tau-\tau_1) \nonumber\\
&=&\int^1_0{\rm d}\tau \int_0^1 \frac{{\rm d}\tau_1}{\tau_1}
M'(\tau_1 W^2) \int_0^1{\rm d}\tau_2 F(\tau_2)\delta (\tau -\tau_1-\tau_2)
\nonumber\\
&=&\int_0^1 \frac{{\rm d}\tau_1}{\tau_1}
M'(\tau_1 W^2) \int_0^1 {\rm d}\tau_2 F(\tau_2)\nonumber\\
&=&\int_0^1 \frac{{\rm d}\tau_1}{\tau_1} M'(\tau_1 W^2)\nonumber\\
&=& <n>_{e^+e^-}\; .
\end{eqnarray}
The final equality in (45) is correct in the accuracy
of the next-to-leading order (see (34) in the last section).
Indeed, the value of $\tau$ cannot exceed 1/2, and therefore the
condition (44) is not satisfied if the multiplicity function $M_g$
is so normalised that the next-to-leading order formula for
$\langle n\rangle_{e^+e^-}$ (34) be satisfied.
Alternatively, we may normalised the multiplicity
function $M_g$ so that the condition (44) be satisfied.
We used the latter normalisation
when we obtained the curve shown in Fig.3.

The condition (44), however, does not completely
remove the error in the normalisation.
As is discussed above, the factorisation formula (41) gives
overestimations of the multiplicity in the large $\tau$ region.
If the multiplicity is normalised by the use of (44), therefore,
the multiplicity is underestimated in the small $\tau$ region.
When we compare the prediction with the experimental data
in the small $\tau$ region,
the error in the normalisation (of the order of $\alpha_s$)
should be taken into account.

\vspace{1.0cm}
We thank M.Yamauchi at KEK for useful discussions.

\newpage
\vspace{2.0cm}
\begin{tabular}
{|c||c|c|c|} \hline
W(GeV) & $\tau$=0.1 & $\tau$=0.2 & $\tau$=0.3 \\ \hline
91 & 1.33 & 1.53 & 1.60 \\ \hline
58 & 1.31 & 1.51 & 1.58 \\ \hline
\end{tabular}

\vspace{5 mm}
\noindent
{\large Table 1} \hspace{1 mm} A comparison of the
predictions of the ratio $\langle n \rangle_{\tau}
/\langle n\rangle_{e^+e^-}$ at different energies.

\vspace{4.0cm}
\noindent {\large \bf FIGURE CAPTIONS}

\vspace{5 mm}
\noindent {\large Fig.1} \hspace{1 mm}
A schematic representation of the
one-particle-inclusive cross section at a given $\tau$.
The dashed line represents the registered particle.
Successive soft-gluon emission takes place in the shaded blob.

\noindent (a) The registered particle originates from the hard quark.

\noindent (b) The registered particle originates from the hard gluon.

\vspace{3 mm}
\noindent {\large Fig.2} \hspace{1 mm}
The prediction for
$(\langle n \rangle_g+\langle n\rangle_{q\bar q})/
\langle n \rangle_{e^+ e^-}$ (solid curve),
and the multiplicity ratio
$R(\tau)=\langle n\rangle_g/\langle n\rangle_q$ (dashed curve).
$W=58$GeV and $\Lambda_{QCD}=0.15$GeV.

\vspace{3 mm}
\noindent {\large Fig.3} \hspace{1 mm}
The predictions for $\langle n\rangle_{\tau}/\langle n\rangle_{e^+e^-}$
($W=58$GeV):
The approximation (41) (solid curve)
vs the approximation (20) (dashed curve).

\end{document}